\documentclass[prb,twocolumn,showpacs,superscriptaddress]{revtex4}

\usepackage{graphicx}
\usepackage{amsmath}
\usepackage{amssymb}
\usepackage{sistyle}
\usepackage{color}

\newcommand{\be}{\begin{equation}}
\newcommand{\ee}{\end{equation}}
\newcommand{\ba}{\begin{eqnarray}}
\newcommand{\ea}{\end{eqnarray}}

\newcount\bozza \bozza=0
\ifnum\bozza=1
\newdimen\shift \shift=-2truecm
\def\lb#1{%
{\label{#1}\rlap{\kern\shift{$\scriptstyle#1$}}}}
\else\def\lb#1{\label{#1}} \fi

\begin{document}

\title{Magneto-optical and optical probes of gapped ground states of bilayer graphene}

\author{E.V. Gorbar}
\affiliation{Department of Physics, Taras Shevchenko National Kiev University, 6 Academician Glushkov ave.,
Kiev 03680, Ukraine}
\affiliation{Bogolyubov Institute for Theoretical Physics, National Academy of Science of Ukraine, 14-b
        Metrologicheskaya Street, Kiev, 03680, Ukraine}

\author{V.P.~Gusynin}
\affiliation{Bogolyubov Institute for Theoretical Physics, National Academy of Science of Ukraine, 14-b
        Metrologicheskaya Street, Kiev, 03680, Ukraine}

\author{A.B.~Kuzmenko}
\affiliation{DPMC, University of Geneva, 1211 Geneva 4, Switzerland}

\author{S.G.~Sharapov}
\affiliation{Bogolyubov Institute for Theoretical Physics, National Academy of Science of Ukraine, 14-b
        Metrologicheskaya Street, Kiev, 03680, Ukraine}

\date{\today }

\begin{abstract}
We study the influence of different kinds of gaps  in a quasiparticle spectrum  on
longitudinal and transverse optical conductivities of bilayer graphene. An exact analytical
expression for magneto-optical conductivity is derived using a low-energy two-band Hamiltonian.
We consider how the layer asymmetry gap caused by a bias electric field and a time-reversal
symmetry breaking gap affect the absorption lines. The limit of zero magnetic field is then
analyzed for an arbitrary carrier density in the two-band model.
For a neutral bilayer graphene, the optical Hall and longitudinal conductivities are calculated
exactly in the four-band model with four different gaps and zero magnetic field.
It is shown that two different time-reversal symmetry breaking states can be
distinguished by analyzing the dependence of the optical Hall conductivity on the energy of photon.
These time-reversal symmetry breaking states are expected to be observed experimentally via optical
polarization rotation either in the Faraday or Kerr effects. We analyze a possibility of such
an experiment for a free-standing graphene, graphene on a thick substrate, and graphene
on a double-layer substrate.
\end{abstract}






\maketitle

\section{Introduction}
\label{sec:intro}

Since its experimental discovery,\cite{Novoselov2006NatPhys} bilayer graphene  became
a separate subject of research due to the features which make it unique among the other
two-dimensional condensed matter systems.
Its low-energy electron spectrum\cite{McCann2006PRL} combines characteristics of
monolayer graphene and traditional two-dimensional electron systems.
It consists of two inequivalent pairs of parabolic valence and conductance bands,
touching each other at $K$ and $K^{\prime}$ points,  and  charge carriers are massive
and possess a chirality.

A unique feature of bilayer graphene is that an electric field $E_\perp$
applied perpendicular to the layers
results in the opening of a tunable gap between the valence and conduction bands.
The value of $E_{\perp}$ can be controlled externally by chemical doping and gating.
Since, in contrast to single layer graphene, the density of states remains finite even
in the unbiased and undoped (neutral) bilayer, there are theoretical predictions%
\cite{Nilsson2006PRL,Zhang2010PRB,Nandkishore1,Nandkishore2010PRL,Vafek2010PRB}
that the electron-electron interaction can result in spontaneous symmetry breaking and
opening a gap even in the absence of a magnetic field.
The nature of the gapped state is much debated in the literature. Possible scenarios include
anomalous quantum Hall (QAH), quantum spin Hall (QSH), layer
antiferromagnet (LAF) states, etc. (see Ref.~\onlinecite{Zhang2011} for a general discussion).
Technically, all these gapped states differ in the way how they break an
approximate $SU(4)$ spin-valley symmetry of the low energy Hamiltonian of bilayer graphene.

In an external magnetic field  symmetric bilayer graphene exhibits anomalous quantum Hall (QH)
effect \cite{Novoselov2006NatPhys} with the filling factors $\nu = \pm 4 n$ with $n=1, 2,\ldots$.
As in single-layer graphene, this QH effect is caused by the anomaly of the zero-energy
lowest Landau level which is eightfold degenerate. The subsequent experiments
\cite{Feldman2009NP,Zhao2010PRL,Weitz2010Science,Martin2010PRL,Kim2011PRL,Freitag2011,Velasco2011,Elferen2011}
showed that in higher magnetic fields the degeneracy of the lowest Landau level is completely resolved and
new QH states with filling factors $\nu=0,\pm1,\pm2,\pm3$ appear. It turned out that the activation energy
gaps for these QH states depend linearly on the  magnetic field $B$. This behavior can be contrasted with
the case of single layer graphene, where the corresponding gaps scale as $\sqrt{B}$.
As suggested in Refs.~\onlinecite{Nandkishore1,Gorbar2010PRB}, the difference between bilayer
and single layer graphene is caused by a strong screening of the Coulomb interaction in the former.

Interestingly, the experiments \cite{Weitz2010Science,Velasco2011}
demonstrated that neutral bilayer graphene remains gapped even when $E_{\perp}=0$ and an applied
magnetic field $B$ vanishes. In particular, the authors of Ref.~\onlinecite{Velasco2011} concluded
that the results of their measurements are most consistent with the LAF state. Nevertheless,
further theoretical and experimental work is necessary to ascertain the nature of the gapped state.

Optical spectroscopy studies proved to be a useful tool in investigation of carbon systems
(see Refs.~\onlinecite{Abergel,Orlita2010SST} for an overview). The relativistic-like gapless
dispersion of quasiparticles in single layer graphene results in a universal and constant
optical conductivity $\sigma_{xx}=e^{2}/4\hbar$ for the photon energies above the threshold
which is twice the Fermi
energy.\cite{Gusynin2006PRL,Falkovsky,Kuzmenko2008PRL,Nair,Basov,Wang,Mak,Dawlaty,Gusynin1009NJP}
Magneto-optical properties of single layer graphene reveal themselves in the spectroscopy
of Landau levels transitions \cite{Sadowski,Kim} and the giant Faraday effect.
\cite{Kuzmenko2011NatPhys}
Optical properties of gapped single layer graphene were intensively studied also in a series of works
both in zero and finite magnetic field.
\cite{Gusynin2006PRB,Gusynin2007JPCM,Morimoto,Koch,Fialkovsky}

Optical methods turned out to be especially fruitful for bilayer graphene.
Theoretical description of its optical
and magneto-optical properties\cite{Abergel2007PRB,Abergel2007EPJ,Nicol2008PRB,Mucha2009JPCM,Mucha2009SSC}
involves more parameters as compared to  single layer graphene and all these parameters were
found experimentally\cite{Zhang2008PRB,Li2009PRL,Kuzmenko2009PRB,Kuzmenko2009aPRB,Mak2009PRL}
in the $B=0$ case as well as in the presence of magnetic field.\cite{Henriksen2008PRL}

As suggested in Ref.~\onlinecite{Levitov2011}, optical methods could also be used for
investigating the symmetry breaking gapped states in bilayer graphene discussed above.
A particular case of the QAH state\cite{Levitov2011} which breaks
time-reversal symmetry explores the idea that such state would show up in the rotation of
the polarization of light.
The papers on magneto-optical conductivity\cite{Abergel2007PRB,Abergel2007EPJ}
do not take into account the presence of gapped states in bilayer graphene, while the
paper\cite{Levitov2011} considers only the case of time-reversal symmetry breaking gapped
state in zero magnetic field. In a recent paper, \cite{Abergel2012} it was shown that
the infrared and far-infrared absorption spectroscopy in bilayer graphene at zero or finite
doping in zero magnetic field can distinguish gapped states from the
gapless unperturbed and nematic states due to their qualitatively different lineshapes.
Since now there is an interest in crossover between various gapped states at finite and zero magnetic field,
in the present paper we derive analytical expressions for magneto-optical conductivity
which include both arbitrary gapped states and finite magnetic field.

The paper is organized as follows. In Sec.~\ref{sec:model}, we introduce the model $4\times 4$ and
$2\times2$ Hamiltonians and discuss various types of gaps. In Sec.~\ref{sec:optical-magnetic}
analytical expressions for optical conductivities in a magnetic field for circularly polarized
light are derived using the two-band model. We discuss how the opening of different kinds
of gaps influences the conductivities. Then, the limit of zero magnetic field is analyzed
for an arbitrary carrier density. In Sec.~\ref{sec:optical-B=0}, we derive the optical Hall
conductivity in zero magnetic field in the presence of time-reversal symmetry breaking gap.
In contrast to Ref.~\onlinecite{Levitov2011}, we use a four-band model to
derive an exact expression for the transverse conductivity which allows us to probe a wider range
of frequencies. We consider the general case where both time-reversal-invariant
and -noninvariant gaps are present. The time-reversal-invariant gap can be
generated due to a perpendicular bias electric field, therefore, we study the
optical Hall conductivity dependence on applied bias electric field.
In Sec.~\ref{sec:Faraday-Kerr}, we consider the detection of
the time-reversal symmetry breaking states in the optical polarization rotation experiment,
viz., by measuring the Faraday or Kerr effects. We analyze a possibility of such an experiment
for a free-standing graphene, graphene on a thick substrate, and graphene on a dielectric layer
on top of a thick layer. Finally, in the Discussion section, we give a brief summary of our results,
and the Appendix at the end of the paper contains an analysis of the optical spectral  weight.

\section{Models and notation}
\label{sec:model}

We consider bilayer graphene in the continuum approximation using both a $4\times 4$
Hamiltonian in the absence of magnetic field $B=0$ and its low-energy $2\times 2$
approximation in the case $B \ne 0$.
The effective Hamiltonian of the four-component model at fixed spin, $s=\pm$, and valley,
$\xi=\pm$, reads as
\be
\label{Hamiltonian-4*4}
H = \xi \left(
\begin{array}{cccc} \Delta_{\xi s} &
0 & 0 & v_F\pi^{\dagger} \\
0 & -\Delta_{\xi s} & v_F\pi & 0 \\
0 & v_F\pi^{\dagger} & 0 & \xi\gamma_1 \\
v_F\pi & 0 & \xi\gamma_1 & 0
\end{array} \right)\,,
\ee
where $\pi=\hat{p}_{x}+i\hat{p}_{y}$, $v_F\approx10^{6}\mbox{m/s}$ is the in-plane Fermi velocity,
$\gamma_1= \SI{0.38}{eV}$ is the inter-layer hopping between pairs of orbitals that lie
directly below and above each other. In our study, we neglect the tight-binding parameters $\gamma_3$
(which leads to a trigonal warping of the band structure at low energies) and $\gamma_4$ whose effects fall
beyond the scope of the present paper.

The Hamiltonian acts on wave function with
four components corresponding to the atomic sites $A1$, $B2$, $A2$, $B1$ in the valley
$K$ ($\xi=+1$) and $B2$, $A1$, $B1$, $A2$ in the valley $K^{\prime}$ ($\xi=-1$).
In the effective Hamiltonian  we included also quasiparticle gaps $\Delta_{\xi s}$
dynamically generated due to the electron-electron interaction.

The possible gapped states \cite{Min2008PRB,Nandkishore2010PRL,Gorbar2010PRB}  can be
classified considering how the gap $\Delta_{\xi s}$ depends on valley and spin indices.
Obviously, selecting gapped states symmetric and antisymmetric in valley and spin, the most
general expression for the gap is given by
\be
\Delta_{\xi s}=U+s U_T+ \xi \Delta_T +\xi s\Delta,
\label{gaps-general}
\ee
where while gaps $U$ and $\Delta$ are invariant with respect to the time
reversal symmetry, gaps $\Delta_T$ and $U_T$ are not.

Among these four gaps the first gap $U$ besides being dynamically generated can be induced by gating
which creates a perpendicular electric field $E_{\perp}$. In this case we will denote it
as $U_0(=eE_{\perp}d/2$), where $d= \SI{0.35}{nm}$ is the distance between layers.
Thus the gap $U$ is related to the layer-polarized state caused by
a potential difference between layers. Note that since the gap term in Eq.~(\ref{Hamiltonian-4*4})
is introduced in the left upper corner, strictly speaking the gap $U$  only approximates the
layer asymmetry gap [see the discussion below Eq.~(\ref{Hamiltonian-2*2})].
In the notations of Ref.~\onlinecite{Zhang2011}, the state related to the gap $U$ is called
quantum valley Hall (QVH) state.
The gap $U_T$ corresponds to the layer antiferromagnet  state,
$\Delta_T$ corresponds  to the quantum anomalous Hall
state,\cite{Nandkishore2010PRL,Nandkishore2010PRB}, and $\Delta$ is related to the spin-polarized
or quantum spin Hall  state.\cite{footnote1}
The corresponding order parameters are summarized in Table~\ref{tab}.
\begin{table*}
\begin{tabular}{|c|c|c|c|}
  \hline
  Ordered & Gap & Order parameter & Broken time-reversal \\
  State      &  &                  & symmetry \\ \hline
  QVH & $U$ & $\langle\psi^{\dagger}_{A1Ks}\psi_{A1Ks}+\psi^{\dagger}_{A1K's}\psi_{A1K's}
-\psi^{\dagger}_{B2Ks}\psi_{B2Ks}-\psi^{\dagger}_{B2K's}\psi_{B2K's}\rangle$ & no \\
  LAF & $U_T$ &  $\langle\psi^{\dagger}_{A1Ks}s\psi_{A1Ks}+\psi^{\dagger}_{A1K's}s\psi_{A1K's}
 -\psi^{\dagger}_{B2Ks}s\psi_{B2Ks}-\psi^{\dagger}_{B2K's}s\psi_{B2K's}\rangle$
  & yes \\
  QAH & $\Delta_T$ & $\langle\psi^{\dagger}_{A1Ks}\psi_{A1Ks}-\psi^{\dagger}_{A1K's}\psi_{A1K's}
 -\psi^{\dagger}_{B2Ks}\psi_{B2Ks}+\psi^{\dagger}_{B2K's}\psi_{B2K's}\rangle$ & yes \\
  QSH & $\Delta$ & $\langle\psi^{\dagger}_{A1Ks}s\psi_{A1Ks}-\psi^{\dagger}_{A1K's}s\psi_{A1K's}
 -\psi^{\dagger}_{B2Ks}s\psi_{B2Ks}+\psi^{\dagger}_{B2K's}s\psi_{B2K's}\rangle$ & no \\
  \hline
\end{tabular}
\caption{Possible gapped states in bilayer graphene at neutral point. QVH, quantum valley Hall; LAF, layer antiferromagnet;
QAH,  quantum anomalous Hall; QSH, quantum spin Hall. The summation over $s=\pm$ is implied.
}
\label{tab}
\end{table*}

All these order parameters except the first one were suggested for describing the ground
state of bilayer graphene at the neutral point in the absence of external electric and magnetic fields.
Indeed, while for a sufficiently large top-bottom voltage difference $U_0$, the layer
polarized QVH state is realized, the experiments \cite{Weitz2010Science,Velasco2011} demonstrated
that as the value $U_0$ decreases there is a phase transition to another state.
This eliminates the QVH state as a possible candidate for the ground state of bilayer
graphene at the neutral point in the absence of external fields.

On the other hand, for sufficiently large magnetic field $B$, the spin-polarized QSH state
is realized. Recent experimental data\cite{Velasco2011}
show the absence of a phase transition as $B$ decreases to zero. This suggests that the QSH
state could be the ground state of bilayer graphene at the neutral point in the absence of external
fields. According to Ref.~\onlinecite{Kharitonov}, the LAF state is adiabatically connected
to the QSH state at high magnetic field, therefore, the LAF state could be also the ground
state of neutral bilayer graphene in the absence of external fields.

To investigate the role of different kinds of gaps in the magneto-transport
properties of bilayer graphene, it is convenient
to use the effective low-energy Hamiltonian which was derived in Ref.~\onlinecite{McCann2006PRL}
using Green's functions. The $2\times 2$ Hamiltonian can also be obtained from
the $4 \times 4$ Hamiltonian
by integrating out $B1$, $A2$  fields which correspond to the Bernal stacked orbitals and
is valid within the energy range $|\epsilon| < \gamma_1/4$. The trigonal warping term
neglected in Eq.~(\ref{Hamiltonian-4*4}) also restricts the validity of the effective Hamiltonian at low
energies.\cite{McCann2006PRL} In an external magnetic field this Hamiltonian takes the form
\begin{equation}
\label{Hamiltonian-2*2}
H^{\mathrm{eff}}=  \left( \begin{array}{cc} \xi\Delta_{\xi s} &
-\frac{(\hat{\pi}^\dagger)^2}{2m} \\
-\frac{\hat{\pi}^2}{2m} & -\xi\Delta_{\xi s} \end{array} \right),
\ee
where  $\hat \pi = \hbar(-i D_x-D_y)$ is now expressed via the covariant derivatives
$D_{i}=\partial_{i}+(ie/\hbar c)A_{i}$ with the electron charge $-e<0$, and the effective mass
of the carriers $m = \gamma_1/(2 v_F^2)$.
The external magnetic field $\mathbf{B} = \nabla \times \mathbf{A}$ is applied perpendicular
to the plane along the positive $z$ axis.
The Hamiltonian (\ref{Hamiltonian-2*2})
acts on a wave function with two components corresponding to the atomic sites $A1$,
$B2$ in the valley $K$ ($\xi=+1$) and $B2$, $A1$ in the valley $K^{\prime}$ ($\xi=-1$).

It has to be noted that different gap terms in the initial $4\times 4$ Hamiltonian
may result in the same effective $2 \times 2$ Hamiltonian. For example, two different
Hamiltonians $H_{U}^{\prime} = \xi \mbox{diag} (U, - U, 0,0)$
which corresponds to Eq.~(\ref{Hamiltonian-4*4}) with $\Delta_{\xi s}=U$ and
$H_{U} = \xi \mbox{diag} (U, - U, - U, U)$ which takes into account the asymmetry
between on-site energies in the two layers,
result in the same $H^{\mathrm{eff}}_{U} = \xi \mbox{diag} (U, - U)$.
One can check that the energy spectrum corresponding to the Hamiltonians
$H_{U}$ and $H_{U}^{\prime}$ is practically identical.
For zero magnetic field the combined effective Hamiltonian with the gap $\Delta_{\xi s}=U$
for two valleys,
$H(\mathbf{p},U)=H^{\mathrm{eff}} (\xi=+1,\mathbf{p},U)\oplus
H^{\mathrm{eff}}(\xi=-1,\mathbf{p},U)$  is time-reversal
invariant under the transformation $(\Pi_{1}\otimes\tau_{1})H^{\ast}(\mathbf{p},U)
(\Pi_{1}\otimes\tau_{1})=H(-\mathbf{p},U)$, where $\Pi_{1}$ swaps $\xi=+1$ and $\xi=-1$
in valley space.\cite{McCann2006PRL} As in the $4 \times 4$ case
the presence of the gap $\Delta_{\xi s}=\xi \Delta_T$ breaks the time-reversal symmetry.
In the presence of a magnetic field the corresponding symmetry transformation becomes $(\Pi_{1}\otimes\tau_{1})
H^{\ast}(\mathbf{p},\Delta_T,B)(\Pi_{1}\otimes\tau_{1})=H(-\mathbf{p},-\Delta_T,-B)$.

We calculate the optical conductivity analytically using the Kubo formula,
\begin{equation}
\label{real-conductivity}
\sigma_{ij}(\Omega)=\frac{\hbar [\Pi_{ij}^R(\Omega+i0)-\Pi_{ij}^R(0)]}{i\Omega},
\end{equation}
where $\Pi_{ij}^R(\Omega+ i0)$ is the retarded current-current correlation function  obtained by analytical
continuation [$\Pi^{R}_{ij}(\Omega )=\Pi_{ij}(i\Omega _{m}\to \Omega +i\epsilon)$]  from its imaginary time
expression, and $\Omega$ is the energy of photon.
Neglecting the  vertex corrections, the calculation of the current-current
correlation function reduces to the evaluation of the bubble diagram
\be
\label{pol_operator}
\begin{split}
&\Pi_{ij}(i\Omega_{m})=-\frac{1}{V}\int\limits_{0}^{\beta }d\tau e^{i\Omega
_{m}\tau }\int d^2rd^2r^\prime \\
& \times {\rm tr}\left[\hat{j_i}({\bf r})G(\mathbf{r},\mathbf{r}',
\tau)\hat{j_j}({\bf r}')G(\mathbf{r}',\mathbf{r},-\tau)\right],
\end{split}
\ee
where $\hat{j_i}({\bf r})=-c\partial H/\partial A^i$ is the electric current density operator,
$G(\mathbf{r},\mathbf{r}',\tau)$ is the electron Green's function (GF),
$V$ is the volume (area) of the system, $\beta = 1/T$ is the inverse temperature, $\Omega_m=
2 \pi m/\beta$, and $\mbox{tr}$ not only takes care of the $4 \times 4$ or $2\times 2$ matrices,
but also includes summation over the valley and spin indices. In
the presence of a magnetic field the GF is not translational invariant  and a special
care should be taken in treating the translation noninvariant phase of the GF as done in
Sec.~\ref{sec:optical-magnetic}. When the magnetic field is absent, the GF's is
translation invariant and one can directly go from Eq.~(\ref{pol_operator}) to the
frequency-momentum representation of the polarization operator as done in Sec.~\ref{sec:optical-B=0}.

\section{Two-band model: optical conductivity in an external magnetic field}
\label{sec:optical-magnetic}

In this section, we will consider the spin singlet gap $\Delta_{\xi}=U+\xi\Delta_T$.
The energies of the Landau levels in the two-band model are
\begin{equation}
\label{LL}
\begin{split}
E_{ n \xi} & =-\xi\Delta_{\xi},\,\, n=0,1, \\
E_{\alpha n \xi} & = \alpha M_{n \xi}, \, M_{n \xi}=\sqrt{\Delta_\xi^{2}+\omega^{2}_{c}n(n-1)},
\, n\ge2,
\end{split}
\end{equation}
where $\alpha=\pm$, $\omega_{c}=\hbar eB/mc=\hbar^{2}/ml^{2} $ is the cyclotron energy, and
$l=\sqrt{\hbar c/eB}$ is the magnetic length.
Using the values $\gamma_1 = \SI{0.38}{eV}$ and $v_F = \SI{1.02e6}{m/s}$
from Ref.~\onlinecite{Kuzmenko2009aPRB} (see also Ref.~\onlinecite{Orlita2010SST}) one can estimate
the effective mass of carriers in bilayer as $m = 0.032 m_e$, where $m_e$ is the electron mass.
Accordingly, the cyclotron energy is equal to $\omega_c \approx \SI{0.116} meV
(m_e/m) B [T] = \SI{3.62} meV B [T] $.

\vspace{5mm}

\subsection{Green's function and calculation of $\Pi_{ij} (\Omega)$}
\label{sec:GF-Pi}

In the two-band model (\ref{Hamiltonian-2*2}) in the Landau gauge $\mathbf{A}=(0,B x)$,
the GF $G(\mathbf{r},\mathbf{r}^{\prime},\omega)$ in the mixed coordinate-frequency
representation has a form
\begin{equation}
G(\mathbf{r},\mathbf{r}^{\prime},\omega)=\exp\left(-i\frac{(x+x')(y-y')}{2l^{2}}\right)
\tilde{G}(\mathbf{r}-\mathbf{r}^{\prime},\omega),
\end{equation}
where the translation invariant part of the GF is represented as a sum over the Landau
levels\cite{Gorbar2010PRB}
\begin{equation}
\label{prop-free-magfield}
\begin{split}
&\tilde{G}(\mathbf{r},\omega)=\frac{1}{2\pi l^{2}}\,
e^{-z/2}\sum\limits_{n=0}^{\infty}\frac{1}{(\omega+\mu)^{2}-M^{2}_{n}}\\
& \times\left(\begin{array}{cc}(\omega+\mu+\xi\Delta_{\xi})L_{n-2}(z)
& \frac{\hbar^{2}(x-i y)^{2}}{2m l^{4}}L^{2}_{n-2}(z) \\
\frac{\hbar^{2}(x+i y)^{2}}{2m l^{4}}L^{2}_{n-2}(z)&(\omega+\mu-\xi\Delta_{\xi})L_{n}(z)
\end{array}\right).
\end{split}
\end{equation}
Here $z=\mathbf{r}^{2}/( 2l^{2})$,  $L_{n}^{\alpha}(z)$ are associated Laguerre polynomials
(by the definition $L^{\alpha}_{-2}(z)=L^{\alpha}_{-1}(z)\equiv0$), and $\mu$ is the chemical potential.
For brevity of notation the subscript $\xi$ in
$M_{n \xi}$ is omitted in what follows, i.e. $M_n=M_{n \xi}$.
Since the $2 \times 2$ Hamiltonian is quadratic in $\hat \pi$ and $\hat \pi^\dagger$,
the electric current operator contains the derivatives
\begin{equation}
\hat{j_x}({\bf r})=\frac{e}{m}\left(\begin{array}{cc}0&\pi^{\dagger}\\
\pi&0\end{array}\right), \quad
\hat{j_y}({\bf r})=\frac{e}{m}
\left(\begin{array}{cc}0&-i\pi^{\dagger}\\
i\pi&0\end{array}\right)\,.
\end{equation}
The phase factors in Eq.~(\ref{pol_operator}) cancel
and we obtain at finite temperature
\begin{equation}
\label{Pi-1}
\begin{split}
\hspace{-2mm}& \Pi_{ij}(i\Omega_{m})=  -\frac{1}{V}\int d^2rd^2r^\prime
 T \sum\limits_{n=-\infty}^{\infty} \\
\hspace{-2mm}& {\rm tr}\left[\tilde{j_i}({\bf r},{\bf r}')\tilde{G}(\mathbf{r},\mathbf{r}',
i\omega_{n})\tilde{j_j}({\bf r}',{\bf r})\tilde{G}(\mathbf{r}',\mathbf{r},i\omega_{n}+i\Omega_{m})\right],
\end{split}
\end{equation}
where $\omega_n = \pi (2n+1)/\beta$ and
\begin{widetext}
\begin{equation}
\begin{split}
\tilde{j_x}({\bf r},{\bf r}')&=\frac{e\hbar}{m}\left(\begin{array}{cc}0&-i\partial_{x}
-\partial_{y}-i\frac{x-x'-i(y-y')}{2l^{2}}\\ -i\partial_{x}
+\partial_{y}+i\frac{x-x'+i(y-y')}{2l^{2}}&0\end{array}\right),\\
\tilde{j_y}({\bf r},{\bf r}')&=\frac{e\hbar}{m}\left(\begin{array}{cc}0&-\partial_{x}
+i\partial_{y}-\frac{x-x'-i(y-y')}{2l^{2}}\\ \partial_{x}
+i\partial_{y}-\frac{x-x'+i(y-y')}{2l^{2}}&0\end{array}\right).
\end{split}
\end{equation}
\end{widetext}
Noting that $\tilde{j_i}({\bf r},{\bf r}')=\tilde{j_i}({\bf r}-{\bf r}')$
depend on the difference of the coordinates ${\bf r}-{\bf r}'$,
one can integrate over ${\bf r}+{\bf r}'$ in Eq.~(\ref{Pi-1}) canceling the volume factor $V$
in the denominator
\begin{equation}
\label{Pi-2}
\begin{split}
& \Pi_{ij}(i\Omega_{m})=T\hspace{-2mm}\sum\limits_{n=-\infty}^{\infty}\int d^2r \\
& {\rm tr}\left[\tilde{j_i}({\bf r})\tilde{G}(\mathbf{r},i\omega_{n})\tilde{j_j}({\bf r})\tilde{G}
(\mathbf{r},i\omega_{n}+i\Omega_{m})\right].
\end{split}
\end{equation}
Writing Eq.~(\ref{Pi-2}) we also used that $\tilde{G}(\mathbf{r},i\omega_{n})$ is even
under the transformation $\mathbf{r}\rightarrow -\mathbf{r}$, while $\tilde{j_i}({\bf r})$
is odd.

The calculation of the optical conductivity follows closely the corresponding calculation done
for monolayer graphene,\cite{Gusynin2006PRB,Gusynin2007JPCM} so we directly proceed to the final expression.
The only difference is that we introduce a finite Landau level width $\Gamma$ at the very end
of the calculation in the final expressions for the conductivities.
The final $\Gamma=0$ expression for the polarization operator
$\Pi_{\pm}^R(\Omega) \equiv \Pi_{xx}^R(\Omega)\pm i\Pi_{xy}^R(\Omega)$
takes the form
\begin{equation}
\label{Pi-Omega-zeroG}
\begin{split}
\Pi_{\pm}^R(\Omega)&=\frac{e^{2}\hbar^{2}}{\pi m^{2}l^{4}}
\sum\limits_{k=0}^{\infty}(k+1)\\
& \times \sum_{\substack{ \lambda,\lambda'=\pm \\
\xi = \pm} }
\frac{n_{F}(\lambda M_{k+1})-n_{F}(\lambda' M_{k+2})}{\lambda' M_{k+2}-\lambda M_{k+1}
\pm(\Omega+i0)}\\
& \times \left[\left(1-\frac{\lambda\lambda'\Delta_{\xi}^{2}}{M_{k+1}M_{k+2}}\right) \pm
\frac{\Omega \lambda \lambda^\prime \xi \Delta_\xi}{M_{k+1}M_{k+2}}
\right],
\end{split}
\end{equation}
where $n_F(\omega) = 1/[\exp((\omega-\mu)/T) +1]$ is the Fermi distribution function.
It is clear that the last $\sim \Omega$ term in $\Pi_{\pm}$  may only be present
if the time-reversal symmetry breaking  gap, $\Delta_T \neq 0$.

\subsection{Magneto-optical conductivity}
\label{sec:magneto-cond}

It is convenient to consider the optical conductivities
$\sigma_{\pm} (\Omega) = \sigma_{xx}(\Omega) \pm i  \sigma_{xy}(\Omega) =\hbar [\Pi_{\pm}^R(\Omega)
-\Pi_{\pm}^R(0)]/(i\Omega )$ which correspond to the opposite circular polarizations of light.
After introducing a finite width of Landau levels in Eq.~(\ref{Pi-Omega-zeroG}),
the final result for the complex conductivities reads as
\begin{equation}
\label{magneto-cond-final}
\begin{split}
 \sigma_{\pm} & (\Omega)=\frac{e^{2}\hbar^{3}}{\pi m^{2}l^{4}}\sum\limits_{k=0}^{\infty}
(k+1)\\
& \times \sum_{\substack{ \lambda,\lambda'=\pm \\ \xi = \pm} }
[n_{F}(\lambda M_{k+1})-n_{F}(\lambda' M_{k+2})]\\
& \times
\left[\frac{1}{\lambda'M_{k+2}-\lambda M_{k+1}}\left(1-\frac{\lambda\lambda'\Delta_\xi^{2}}
{M_{k+1}M_{k+2}}\right) \right. \pm \\
& \left.  \frac{\lambda \lambda^\prime \xi \Delta_\xi}{M_{k+1}M_{k+2}} \right]
\frac{i}{\Omega\mp\lambda' M_{k+2}\pm\lambda M_{k+1}+2 i \Gamma}.
\end{split}
\end{equation}
The expressions for conductivities in the case of more general gaps $\Delta_{\xi s}$
are obtained from Eq.~(\ref{magneto-cond-final}) by replacing $\Delta_{\xi}$ with $\Delta_{\xi s}$
and inserting the overall factor $1/2$ and summing over the spin variable.
Equation~(\ref{magneto-cond-final}) is one of the main results of this paper which generalizes
the results of Refs.~\onlinecite{Abergel2007PRB,Abergel2007EPJ} to a finite $\Delta_\xi$ case.
For $\Delta_\xi =0$, there is a practically overall agreement between Eq.~(\ref{magneto-cond-final}) and
the corresponding expressions from Ref.~\onlinecite{Abergel2007PRB,Abergel2007EPJ} except to the intensity of the
$0$ to $1$ transition.
To verify this issue, in Appendix~\ref{sec:spectral-weight} we consider the behavior of
the spectral weight. We show both that our result agrees with the behavior of the spectral weight for $B=0$
and that the weight is conserved when the chemical potential moves from the region
$M_0 < \mu < M_2$ to the region $\mu > M_2$.

It has to be stressed that rather simple analytical result (\ref{magneto-cond-final}) was possible to obtain
because we used the effective $2 \times 2$ Hamiltonian. The $4 \times 4$ consideration is much more involved
\cite{Mucha2009JPCM,Mucha2009SSC}. Moreover, the neglected tight-binding terms, including the trigonal warping,
can only be treated numerically\cite{footnote} which is beyond the scope of the present work.

The scheme of Landau levels and allowed transitions for bilayer can be found, for example, in the
review,\cite{Orlita2010SST} so that we can go directly to the discussion of the behavior of magneto-conductivity.
In Figs.~\ref{fig:1} and \ref{fig:2} we show the results based on the computation of Eq.~(\ref{magneto-cond-final})
for the real part of $\sigma_{\pm}(\Omega)$ in units of $\sigma_0= e^2/(4\hbar)$ as a function of $\Omega$
measured in meV. The highest value of $\Omega = \SI{80}{mev}$ is less than $\gamma_1/4$.
In both figures we set $B= \SI{3}{T}$, $T=\SI{5}{K}$ and $\Gamma = \SI{2}{meV}$.

The reference case\cite{Abergel2007PRB,Abergel2007EPJ} $\Delta_\xi=0$ is shown in Fig.~\ref{fig:1}.
To scrutinize the curves in this figure, we provide the energies of the Landau levels, viz.
for the $n=2$ level
$|E_{\pm 2}| = M_2 \approx \SI{15.3}{meV}$, for the $n=3$ level
$|E_{\pm 3}| = M_3 \approx \SI{26.6}{meV}$ and  for the $n=4$ level
$|E_{\pm 4}| = M_4 \approx \SI{37.6}{meV}$.
The long dashed (red) curve is for $\sigma_{+}(\Omega)$ and the dash-dotted (black) curve
is for $\sigma_{-}(\Omega)$. Both curves are for $\mu = \SI{10}{meV}$ which is less than $E_2$.
Accordingly, in the long dashed (red) curve we observe the lines with the energy
$M_2 - M_{0,1} \approx \SI{15.3}{meV}$, $M_2 + M_3 \approx \SI{41.9}{meV}$ and $M_3 + M_4
\approx \SI{64.2}{meV}$.
They correspond to the transitions from the Landau levels with the energy $E_{0,1}$ to $E_2$,
from $E_{-2}$ to $E_3$, and from $E_{-3}$ to $E_4$, respectively. Since
for the opposite polarization of light presented by the dash-dotted (black) curve there is no transition
between the 0th ($n=0,1$) and $E_2$ level, the first observed line is for the transition
from $E_{-3}$ to $E_2$ levels
with the energy $M_2+ M_3$ coinciding with the second line on the previous curve.
The second observed line corresponds to the transition from $E_{-4}$ to $E_3$ levels with
the energy $M_3 + M_4$.
The solid (blue) curve is for $\sigma_{+}(\Omega)$ and the short dashed (green) curve
is for $\sigma_{-}(\Omega)$, and both curves are for $\mu = \SI{20}{meV}$, so that $E_{2} < \mu < E_3 $.
Accordingly, in the solid (blue) curve the transition with the energy $M_2 - M_1$ is impossible,
but instead the transition with the energy $M_3 - M_2 \approx \SI{11.3}{meV}$ develops.
Two remaining transitions from $E_{-2}$ to $E_3$, and from $E_{-3}$ to $E_4$
which coincide with the transitions on the considered
above long dashed (red) and dash-dotted (black) curves are also present.
For the opposite polarization of light presented by the short dashed (green) curve
the only possible transition is from $E_{-4}$ to $E_3$, so that the peak at the energy $M_3 + M_4$
is the only peak shared by all curves.
\begin{figure}[ht]
\centering{
\includegraphics[width=8cm]{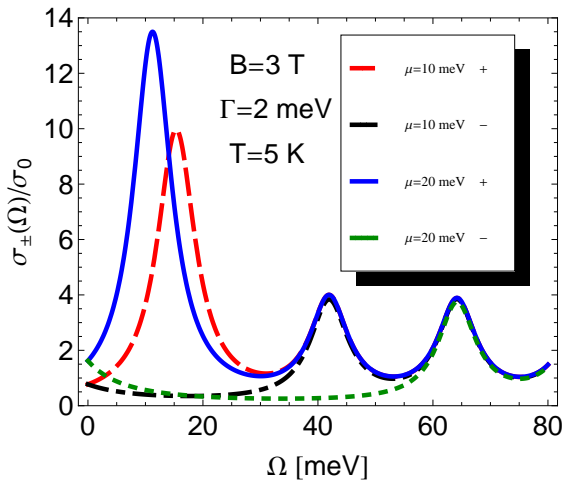}
}
\caption{(Color online) The real part of the conductivity $\sigma_{\pm}(\Omega)$
in units of $\sigma_0= e^2/(4\hbar)$ versus the photon energy $\Omega$
in meV for magnetic field $B= \SI{3}{T}$, temperature $T=\SI{5}{K}$, and
scattering rate $\Gamma = \SI{2}{meV}$. Long dashed ($\sigma_{+}$)
and dash-dotted ($\sigma_{-}$) are for the chemical potential $\mu = \SI{10}{meV}$. The solid
($\sigma_{+}$) and short dashed ($\sigma_{-}$)  are for the chemical potential $\mu = \SI{20}{meV}$.}
\label{fig:1}
\end{figure}

\subsection{Comparison of the QVH and QAH states}
\label{sec:QVH-QAH}

Now we are at the position to discuss how the opening of either time-reversal symmetry breaking gap $\Delta_T$
or preserving this symmetry gap $U$ changes the presented above picture
(see Fig.~\ref{fig:2}).
\begin{figure}[ht]
\centering{
\includegraphics[width=8cm]{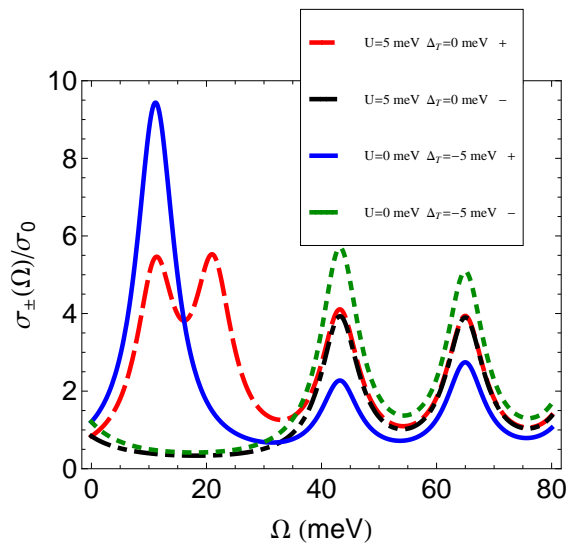}
}
\caption{(Color online) The real part of the conductivity $\sigma_{\pm}(\Omega)$
in units of $\sigma_0= e^2/(4\hbar)$ versus the photon energy $\Omega$
in meV for magnetic field $B= \SI{3}{T}$, temperature $T=\SI{5}{K}$,
scattering rate $\Gamma = \SI{2}{meV}$, and the chemical potential $\mu = \SI{10}{meV}$.
Long dashed ($\sigma_{+}$) and dash-dotted ($\sigma_{-}$) are for the layer asymmetry gap
$U = \SI{5}{meV}$ and the QAH gap $\Delta_{T} =0$.
The solid ($\sigma_{+}$) and short dashed ($\sigma_{-}$)  are for the layer asymmetry gap
$U = 0$ and the QAH gap $\Delta_{T} =\SI{5}{meV}$.
}
\label{fig:2}
\end{figure}

Again long dashed (red) curve is for $\sigma_{+}(\Omega)$ and the dash-dotted (black) curve
is for $\sigma_{-}(\Omega)$. Both curves are for $\mu = \SI{10}{meV}$, $U =
\SI{5}{meV}$ and $\Delta_T=0$. The solid (blue) curve is for $\sigma_{+}(\Omega)$ and the
short dashed (green) curve is for $\sigma_{-}(\Omega)$, and while the value of
the chemical potential is kept the same, $\mu = \SI{10}{meV}$, the gaps are now
$U = 0$ and $\Delta_T =\SI{5}{meV}$.
For relatively small values of the gaps, $\SI{5}{meV}$, the energies of the levels
with $n\neq 0,1$ remain
practically intact, viz. $|E_{\pm 2}| = M_2 \approx \SI{16.1}{meV}$,
$|E_{\pm 3}| = M_3 \approx \SI{27.1}{meV}$ and
$|E_{\pm 4}| = M_4 \approx \SI{37.9}{meV}$.
Accordingly, the last two peaks from the right remain practically unchanged, the
first from the right peak with the
energy $M_3+M_4$ shifts in energy from $\SI{64.2}{meV}$ to $\SI{65}{meV}$, and the
second from the right peak
 with the energy $M_2+M_3$ shifts in energy from $\SI{41.9}{meV}$ to $\SI{43.2}{meV}$. The
most essential changes occur due to the partial removal of the degeneracy of the lowest Landau level.

Since we took $\mu = \SI{10}{meV}$ all curves in Fig.~\ref{fig:2} have to be compared with
the  two curves in Fig.~\ref{fig:1} plotted for the same value of the chemical potential, viz.
the long dashed $\sigma_{+}(\Omega)$ curve and the dash-dotted $\sigma_{-}(\Omega)$ curve.

Let us begin with the $\sigma_{+}(\Omega)$ polarization.
We observe that for the gap $U = \SI{5}{meV}$
the peak at $\SI{15.3}{meV}$ corresponding to the transition from
$E_{0,1}$ to $E_2$ levels is split into two peaks at $\SI{11.1}{meV}$ and
$\SI{21.1}{meV}$. These new peaks correspond to the transitions from the levels
$E_{0,1} =  U$ and $E_{0,-1} = - U$
to $E_2$ level, respectively. For the time-reversal symmetry breaking gap,
$\Delta_T = \SI{5}{meV}$, we observe that there is only one peak at $\SI{21.1}{meV}$ which corresponds
to the transition from the only level $E_{0,-1} = - \Delta_T$ to $E_2$.
One can verify that the position of the peak is sensitive to the sign of the gap, i.e.
for $\Delta_T = -\SI{5}{meV}$ the position of the peak is at $\SI{11.1}{meV}$, because
there is the only level at the positive energy, $E_{0,1} = - \Delta_T>0$.

For the $\sigma_{-}(\Omega)$ polarization, the discussed above transitions are forbidden.
The only small difference is observed between the height of the second from the right
peaks in the solid (blue) and short dashed (green) curves. These curves interchange when
the sign of the gap $\Delta_T$ is reversed to $-\Delta_T$.

It should be noted that although a measurement of the optical Hall conductivity $\sigma_{xy}(\Omega)$
allows one to distinguish holes and electrons,\cite{Crassee2011PRB} some essential features such as
splitting of the absorption peak into  two peaks depending on the type of the gap
remain present even in the diagonal conductivity $\sigma_{xx}(\Omega)$.

The obtained results allow us to conclude that the investigation of the optical magneto-conductivity
may provide an additional insight into the nature of the gaps in the bilayer graphene.
The features discussed here could be observed if the condition $\Gamma < |\Delta_{\xi}|$ is satisfied.
Since our value $\Gamma = \SI{2}{meV}$ is almost 5 times smaller than that used in
Ref.~\onlinecite{Kuzmenko2009aPRB} for fitting the data on the gated bilayer at $B=0$,
the observation of the discussed here effects requires samples of a much better quality.

\subsection{Zero-field and dc limits}
\label{sec:low-B}

Starting from Eq.~(\ref{magneto-cond-final}) one can reproduce correctly $B\to 0$
limit. Similarly to the consideration done for monolayer graphene,\cite{Gusynin2007JPCM} introducing
a continuum variable $\omega$ instead of $M_n$ given by Eq.~(\ref{LL}) and replacing the sum over $n$
by the integral, in the limit $\Gamma\to 0$ for the real part of the diagonal optical conductivity
$\sigma_{xx}(\Omega) = (\sigma_{+}(\Omega) +  \sigma_{-}(\Omega))/2$
one obtains
\begin{equation}
\begin{split}
{\rm Re}\, & \sigma_{xx}(\Omega)=
\frac{\pi e^{2}}{2h}\sum\limits_{\xi,s=\pm}\left\{
2\delta(\Omega)\int\limits_{|\Delta_{\xi s}|}^{\infty}d\omega \omega \left(1-\frac{\Delta_{\xi s}^{2}}{\omega^{2}}
\right) \right.\\
& \times \left.
\left(\frac{1}{4T\cosh^{2}\frac{\omega-\mu}{T}}+\frac{1}{4T\cosh^{2}\frac{\omega+\mu}{T}}\right)\right.\\
&+\left.\frac{1}{2}\left(1+\frac{4\Delta_{\xi s}^{2}}{\Omega^{2}}\right)
\frac{\sinh(|\Omega|/2T)\theta(|\Omega|-2|\Delta_{\xi s}|)}{\cosh(\mu/T)+\cosh(\Omega/2T)}\right\}.
\end{split}
\end{equation}
Here we also included the dependence of the gap $\Delta_{\xi s}$ on the spin variable $s$.
The first term in curved brackets corresponds to the Drude peak while the second one
describes interband electron-photon scattering processes.

Finally, at $T=0$ we obtain
\begin{equation}
\label{AC-B=T=0}
\begin{split}
{\rm Re}\,\sigma_{xx}(\Omega)=&\frac{\pi e^{2}}{h}\sum\limits_{\xi,s=\pm}\left[\delta(\Omega)
\frac{\mu^{2}-\Delta_{\xi s}^{2}}{|\mu|}\theta(\left|\mu|-|\Delta_{\xi s}|\right) \right.\\
&+ \left. \frac{\Omega^{2}+4\Delta_{\xi s}^{2}}{4\Omega^{2}}\theta(|\Omega|-2\mbox{max}(|\mu|,
|\Delta_{\xi s}|)\right].
\end{split}
\end{equation}
This expression for longitudinal optical conductivity is similar to that obtained
for single layer graphene\cite{Gusynin2006PRL} and for topological insulators.\cite{Tse-PRL}
It is worth noting that there is a more deep analogy between the band structures and
optical conductivities of the single layer graphene with Rashba term and biased bilayer
graphene.\cite{Ingenhoven2010PRB}

Similarly, one can consider the $B\to 0$ limit for the optical Hall conductivity,
$\sigma_{xy}(\Omega) = (\sigma_{+}(\Omega) -  \sigma_{-}(\Omega))/(2i)$ which takes the form
\begin{equation}
\sigma_{xy}(\Omega)=-\frac{4e^{2}}{h}\sum\limits_{\xi,s=\pm}\xi\Delta_{\xi s}\int
\limits_{|\Delta_{\xi s}|}^{\infty}d\omega\frac{n_{F}(\omega)-n_{F}(-\omega)}{4\omega^2-(\Omega+i0)^{2}}.
\ee
Note that the last expression is an even function of chemical potential $\mu$ in contrast
to the Hall conductivity in a magnetic field which is an odd function of $\mu$.
It is clear that  $\sigma_{xy}(\Omega) \neq 0$ only if the time-reversal
symmetry breaking gaps $U_{T}$ or $\Delta_{T}$ are nonzero. For zero temperature we obtain
\begin{equation}
\begin{split}
\sigma_{xy}(\Omega)&=\frac{e^{2}}{h(\Omega+i0)}\sum\limits_{\xi,s=\pm}\xi\Delta_{\xi s}\\
&\times\ln\frac{2\mbox{max}(|\mu|,|\Delta_{\xi s}|)+\Omega+i0}{2\mbox{max}(|\mu|,|\Delta_{\xi s}|)-(\Omega+i0)}.
\label{sigmaxy-zeroT}
\end{split}
\end{equation}
Equation~(\ref{sigmaxy-zeroT}) resembles the result obtained for topological insulators. \cite{Tse-PRL}
Making the change of the variable, $s\rightarrow\xi s$, in Eqs.(\ref{AC-B=T=0}) and (\ref{sigmaxy-zeroT}),
one can see that both conductivities are invariant under the interchange $U_{T}\leftrightarrow
\Delta.$
The real part of the dc Hall conductivity takes the form
\begin{equation}
\label{dc-Hall-1}
\begin{split}
{\rm Re}\,\sigma_{xy}&=\frac{e^{2}}{h}\sum_{\xi,s=\pm}\xi\frac{\Delta_{\xi s}}{\mbox{max}
(|\mu|,|\Delta_{\xi s}|)}\\
& =
\frac{e^{2}}{h}\sum_{\xi,s=\pm}\xi \left\{
\begin{array}{cc}
{\rm sgn}(\Delta_{\xi s}), & |\mu|<|\Delta_{\xi s}|,\\
\frac{\Delta_{\xi s}}{|\mu|}, & |\mu| \geq |\Delta_{\xi s}|,
\end{array}%
\right.
\end{split}
\end{equation}
which  at $\mu=0$, $U=U_{T}=\Delta=0$, and $\Delta_T \neq0$ is in agreement with the corresponding
expression in Ref.~\onlinecite{Tse-PRB}. At the neutral point, $\mu=0$, we have,
\begin{equation}
{\rm Re}\,\sigma_{xy}=\frac{\nu e^{2}}{h},\quad \nu=\sum\limits_{\xi,s=\pm}\xi\,{\rm sgn}(\Delta_{\xi s}).
\label{Hall-conductivity}
\end{equation}
Clearly, the factor $\nu$ can take the values $\nu=0,\pm2,\pm4$ depending on the relations
between the gaps $U,U_T,\Delta_T, \Delta$ and their signs.
The QAH gap $\Delta_T$ and LAF gap $U_T$ break time-reversal
symmetry and this is the necessary condition for observation of nonzero Hall conductivity.
For example, $\nu=4$ is realized in the case $\Delta_T>0,U=U_T=
\Delta=0$ (for $\Delta_T<0$, obviously, $\nu=-4$), thus QAH phase has a zero-field
quantized charge Hall conductivity. In the QSH phase with $\Delta\neq0$
and $U=U_{T}=\Delta_{T}=0$, two spin components have
opposite Hall conductivity, hence zero charge Hall conductivity. On the other hand, in this phase a spin Hall
conductivity $\sigma_{xy}(s=+)-\sigma_{xy}(s=-)$ is nonzero and quantized.
For $\Delta_T=0$ and taking without the loss of generality all other gaps positive, we obtain
$\nu=-2$ if the following conditions are satisfied
\begin{equation}
\begin{split}
&U+U_T>\Delta>|U-U_T|,\\
&U+{\Delta}>U_{T}>|U-{\Delta}|
\label{inequalities}
\end{split}
\end{equation}
[$\nu=2$ is obtained if we invert the signs of all three gaps and replace the gap values by their
absolute values]. The second inequality in (\ref{inequalities}) arises as a consequence
of the symmetry $\sigma_{xy}(\Omega)$ under the interchange $U_{T}\leftrightarrow {\Delta}$.

In Fig.~\ref{fig:3} we plotted the real part of frequency dependent Hall
conductivity at zero temperature and scattering rate $2 \Gamma = \SI{5}{meV}$. The scattering rate $\Gamma$
is introduced by replacing $i0$ in Eq.(\ref{sigmaxy-zeroT}) by  $2 i\Gamma$.
In Fig.~\ref{fig:3}
two different cases for time-reversal breaking gaps are presented: 1) Long dashed (red) line is
for a finite QAH gap $\Delta_{T}= \SI{5}{mev}$, all the other gaps are zero, $U=U_{T}=\Delta=0$,
and 2) For solid (blue)
line the gaps are taken $\Delta_{T}=0$, $U= \SI{5}{mev}$, $U_{T}=\SI{4}{mev},
\Delta=\SI{1.5}{mev}$. Since the two-band model is valid for $\Omega \leq  \gamma_{1}/4\sim \SI{0.1}{eV}$,
only this range of energies is shown in the figure.
\begin{figure}[ht]
\centering{
\includegraphics[width=8cm]{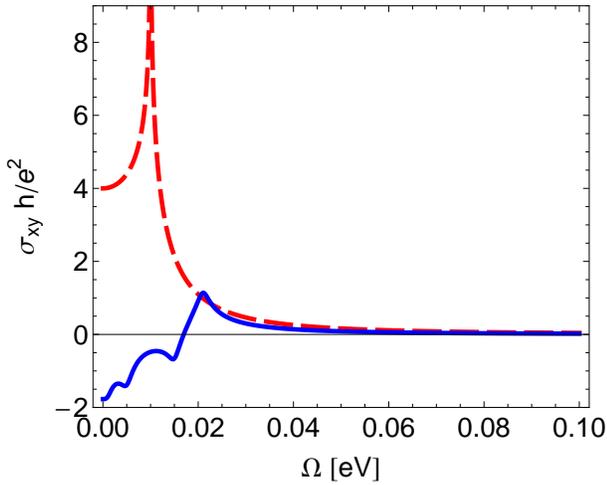}
}
\caption{ (Color online) The real part of the optical Hall conductivity $\sigma_{xy}(\Omega)$
in units of $e^2/h$ versus the photon energy $\Omega$ in eV for $T = \mu =0$, the scattering rate
$2 \Gamma = \SI{5}{meV}$. The long dashed (red) line is for the QAH gap $\Delta_{T}= \SI{5}{mev}$,
the other gaps are zero, $U=U_{T}=\Delta=0$. The solid line is for $\Delta_{T}=0$, the layer
asymmetry gap $U=\SI{5}{mev}$, the LAF gap $U_{T}=\SI{4}{mev}$, and the QSH gap
$\Delta= \SI{1.5}{mev}$.}
\label{fig:3}
\end{figure}

These two types of time-reversal symmetry breaking gaps in bilayer
graphene are expected to manifest themselves through a nonzero dc Hall response in the absence
of a magnetic field. At finite frequencies, the behavior of the Hall conductivities differs
essentially for two types of gaps: the curve for nonzero gap $\Delta$ crosses zero, while the curve for
$\Delta_{T}\neq0$ does not. The crossing takes place in infrared range of frequencies, while
at  near-infrared and optical frequencies, $\Omega\sim\gamma_{1}/4\gg|\Delta_{\xi s}|$, the Hall conductivity
is small and of order ${\rm Re}\sigma_{xy}\sim \sum_{\xi,s}\Delta_{\xi s}|\Delta_{\xi s}|/\Omega^{2}$
[see, Eq.~(\ref{sigmaxy-zeroT})].
Nevertheless, it is possible to observe such  gaps in a Hall response at optical frequencies
as was suggested in Ref.~\onlinecite{Levitov2011}. The point is that the Hall conductivity at
optical frequencies is dominated by transitions to the high-energy bands and can be manifested
in the polar Kerr or Faraday rotation. However, to make reliable calculations, it is necessary 
to move up to the four-band model. We study the optical Hall conductivity in the four-band model in Sec.~\ref{sec:optical-B=0}.

\section{Optical Hall conductivity in the four-component model for zero magnetic field}
\label{sec:optical-B=0}

In this section, we will calculate and analyze the optical Hall conductivity
for bilayer graphene at the neutral point, i.e. $\mu =0$ is set from the beginning.
It was shown Ref.~\onlinecite{Levitov2011} that if $\Delta_T \neq 0$, the ac Hall conductivity, $\sigma_{xy}(\Omega)$
exhibits a resonant enhancement at $\Omega = \gamma_1$ which corresponds to the optical frequencies
due to transitions from the low-energy bands to the high-energy bands.
These optical interband transitions were effectively introduced into the two-band model\cite{Levitov2011}
using projector operators.

Here, instead, we use the full four-band model (\ref{Hamiltonian-4*4}) and obtain an exact
expression for optical Hall conductivity valid in a wide range of the photon energies.
This enables us to investigate how behavior of $\sigma_{xy}$ is affected by the other
types of the gaps both in the infrared and optical ranges of frequencies.

In the four-band model, the GF for quasiparticles at fixed spin and valley equals
\begin{equation}
\begin{split}
G(\omega,\mathbf{p}) =(\omega & -H)^{-1}=\\
\frac{1}{(\omega^2-E^2_1(\mathbf{p}))(\omega^2-E^2_2(\mathbf{p}))} & \left(
\begin{array}{cc} A_1 &
A_2 \\
A_3 & A_4 \end{array} \right)\,,
\label{GF-Levitov}
\end{split}
\end{equation}
where
\begin{equation}
\begin{split}
E^2_{1,2}(\mathbf{p})& =\frac{\gamma_1^2+\Delta_{\xi s}^{2}}{2}+v_F^2\mathbf{p}^2 \\
& \pm \sqrt{\frac{(\gamma_1^2-\Delta_{\xi s}^{2})^{2}}{4}+v_F^2\mathbf{p}^2(\gamma^{2}_{1}+\Delta_{\xi s}^2)},
\end{split}
\label{energy-spectrum}
\end{equation}
and the block matrices are
\begin{widetext}
\begin{equation}
\begin{split}
&A_{1}=\left(\begin{array}{cc}
(\omega+\xi\Delta_{\xi s})(\omega^2-\gamma_1^2)-\omega v^{2}_{F}\mathbf{p^2}
&
v^{2}_{F}\gamma_1(p_{x}-i p_{y})^2 \\
v^{2}_{F}\gamma_1(p_{x}+i p_{y})^2&
(\omega-\xi\Delta_{\xi s})(\omega^2-\gamma_1^2)-\omega v^{2}_{F}\mathbf{p^2}
\end{array} \right),\\
&A_{2}=\xi\left(\begin{array}{cc}
 v_{F}\gamma_1(\omega+\xi\Delta_{\xi s})(p_{x}-i p_{y}) &  v_{F}[\omega(\omega+\xi\Delta_{\xi s})
-v^{2}_{F}\mathbf{p}^2](p_{x}-i p_{y}) \\
 v_{F}\,[\omega(\omega-\xi\Delta_{\xi s})-v^{2}_{F}\mathbf{p}^2](p_{x}+i p_{y}) &  v_{F}\gamma_1
(\omega-\xi\Delta_{\xi s})(p_{x}+i p_{y})
\end{array} \right),\\
&A_{3}=\xi\left(\begin{array}{cc}
 v_{F}\gamma_1(\omega+\xi\Delta_{\xi s})(p_{x}+i p_{y}) & v_{F}[\omega(\omega-\xi\Delta_{\xi s})-
 v^{2}_{F}\mathbf{p}^2](p_{x}-i p_{y}) \\
v_{F}[\omega(\omega+\xi\Delta_{\xi s})-v^{2}_{F}\mathbf{p}^2](p_{x}+i p_{y}) & v_{F}
\gamma_1(\omega-\xi\Delta_{\xi s})(p_{x}-i p_{y}),
\end{array} \right),\\
&A_{4}=\left(
\begin{array}{cc}
(\omega+\xi\Delta_{\xi})(\omega(\omega-\xi\Delta_{\xi s})-v^{2}_{F}\mathbf{p^2}) &
\gamma_1(\omega^2-\Delta_{\xi s}^2) \\
\gamma_1(\omega^2-\Delta_{\xi s}^2) &
(\omega-\xi\Delta_{\xi s})(\omega(\omega+\xi\Delta_{\xi s})-v^{2}_{F}\mathbf{p^2})
\end{array} \right).
\end{split}
\end{equation}
\end{widetext}
For $|\Delta_{\xi s}|\ll v_Fp\ll\gamma_1$, we find for positive energies,
\ba
E_{1}(\mathbf{p})\simeq \gamma_{1}+\frac{v_F^2\mathbf{p}^2}{\gamma_1},
\quad E_{2}(\mathbf{p})\simeq \sqrt{|\Delta_{\xi s}|^2+\frac{v_F^4\mathbf{p}^4}{\gamma_1^2}}\,,
\label{energies-infrared}
\ea
i.e., $E_1(\mathbf{p})$ and $E_2(\mathbf{p})$ describe the positive energy branches of
the high and low energy bands, respectively (the negative energy branches of these bands
are obtained multiplying the above expressions by $-1$).

According to the Kubo formula, the tensor of conductivities is given by Eq.~(\ref{real-conductivity}),
where at zero temperature
\be
\Pi_{ij}^R (\Omega+i0)=i\int
\frac{d\omega d^{2}p}{(2\pi )^{3}}{\rm tr}\left[ j_{i}G(\omega,{\mathbf{p}})j_{j}
G(\omega-\Omega,{\mathbf{p}})\right]\,,
\label{sigma_spectral_repr}
\ee
and the trace includes the summation over spin and valley degrees of freedom. Here, the current density
$\mathbf{j}=e \partial H/\partial\mathbf{p}$,
or in components,
\be
j_{x}=e\xi v_{F}\left(
\begin{array}{cccc}0&0&0&1\\ 0&0&1&0\\ 0&1&0&0\\ 1&0&0&0\end{array} \right),\quad
j_{y}=e\xi v_{F}\left(
\begin{array}{cccc}0&0&0&-i\\ 0&0&i&0\\ 0&-i&0&0\\ i&0&0&0\end{array} \right).
\ee
Using Eq.~(\ref{sigma_spectral_repr}) and taking the trace, we get
\begin{equation}
\Pi_{xy}^R(\Omega)=-\frac{2 e^{2}v_F^2\Omega}{\hbar^{2}}\sum\limits_{\xi=\pm, s=\pm}\hspace{-2mm}
\xi\Delta_{\xi s}\int\frac{d\omega d^{2}p}
{(2\pi)^{3}}\frac{N(\omega,\mathbf{p})}{D(\omega,\mathbf{p})},
\label{pol-function-Omega}
\end{equation}
where the numerator and denominator are, respectively,
\begin{equation}
\begin{split}
{N(\omega,\mathbf{p})}&=v_F^4\mathbf{p}^{4}+v_F^2\mathbf{p}^{2}\left(4\gamma_{1}^{2}
-\omega^{2}-(\omega-\Omega)^{2}\right)\\
&+\omega^{2}(\omega-\Omega)^{2} +\Delta_{\xi s}^{2}\omega(\omega-\Omega)\\
&+\gamma_{1}^{2}\left(\Delta_{\xi s}^{2}-3\omega(\omega-\Omega)-\Omega^{2}\right),\\
D(\omega,\mathbf{p})&=
[\omega^2-E^2_1(\mathbf{p})][\omega^2-E^2_2(\mathbf{p})]\\
& \times [(\omega-\Omega)^2-E^2_1(\mathbf{p})]
[(\omega-\Omega)^2-E^2_2(\mathbf{p})]\,.
\end{split}
\end{equation}
Eq.~(\ref{pol-function-Omega}) implies that the function
$\Pi_{xy}^R(\Omega)$ is an odd function of the energy: $\Pi_{xy}^R(-\Omega)=-\Pi_{xy}^R(\Omega)$.
On the other hand, from Eq.~(\ref{sigma_spectral_repr}) we obtain that
$\Pi_{yx}^R(\Omega)=\Pi_{xy}^R(-\Omega)$, therefore, for the Hall conductivity we have the relationship
$\sigma_{yx}(\Omega)=-\sigma_{xy}(\Omega)$.

The integration over $\omega$ and the angle in Eq.~(\ref{pol-function-Omega}) can be done
straightforwardly and we obtain the final general expression for the optical Hall conductivity
in the four-component model of bilayer graphene,
\begin{widetext}
\begin{equation}
\begin{split}
\sigma_{xy}(\Omega)&=\frac{e^{2}}{2h} \sum\limits_{\xi=\pm, s=\pm}
\xi\Delta_{\xi s}\int\limits_{0}^{\infty}dx\left\{\frac{2\gamma_{1}^{2}x}
{(E_{1}^{2}-E_{2}^{2})^{2}}\left[\frac{1}{E_{1}^{2}}
\left(\frac{1}{2E_{1}-\Omega-2i\Gamma}+\frac{1}{2E_{1}+\Omega+2i\Gamma}\right)\right.\right.\\
&+\left.\left.\frac{1}{E_{2}^{2}}
\left(\frac{1}{2E_{2}-\Omega-2i\Gamma}+\frac{1}{2E_{2}+\Omega+2i\Gamma}\right)\right]\right.\\
&+ \left.\frac{4\gamma_{1}^{4}-\gamma_{1}^{2}(3E_{1}^{2}+2E_{1}E_{2}+3E_{2}^{2}-
4\Delta_{\xi s}^{2})+(E_{1}^{2}-E_{2}^{2})^{2}-(E_{1}-E_{2})^{2}\Delta_{\xi s}^{2}}
{2E_{1}E_{2}(E_{1}^{2}-E_{2}^{2})^{2}}\right.\\
&\times\left.\left(\frac{1}{E_{1}+E_{2}-\Omega-2i\Gamma}
+\frac{1}{E_{1}+E_{2}+\Omega+2i\Gamma}\right)\right\}.
\label{sigmaxy-clean}
\end{split}
\end{equation}
\end{widetext}
Here as before we introduced a phenomenological impurity scattering rate $\Gamma$.
In this form, the physical meaning of three terms in Eqs.~(\ref{sigmaxy-clean})
is quite transparent. Clearly, the first term  in the square brackets
describes transitions between the negative and positive branches of the high energy band
$E_1(\mathbf{p})$, the second term is related to transitions between the negative and positive
branches of the low energy band $E_2(\mathbf{p})$, and the last term in curly brackets
describes interband transitions.

The dc Hall conductivity in clean sample is obtained from Eq.~(\ref{sigmaxy-clean})
setting $\Gamma=0$ and $\Omega=0$. It can be rewritten
in terms of the dimensionless variables $y=x/\gamma^{2}_{1}$, $z=\Delta_T/\gamma_{1}$
($2y=M_{1}^{2}+M_{2}^{2}-1-z^{2}$) and for the time-reversal symmetry breaking gap,
$\Delta_{\xi s}=\xi\Delta_T$,  it acquires the form:
\begin{equation}
\begin{split}
\sigma_{xy}&=\frac{4e^{2}z}{h}\int\limits_{0}^{\infty}dy\left[\frac{1+M_{1}M_{2}}
{M_{1}M_{2}(M_{1}+M_{2})^{3}}\right.\\
&+\left.y\frac{3M_{1}M_{2}+M_{1}^{2}+M_{2}^{2}+M_{1}^{2}M_{2}^{2}}
{M_{1}^{3}M_{2}^{3}(M_{1}+M_{2})^{3}}\right],
\label{dc-Hall}
\end{split}
\end{equation}
where
\ba
\hspace{-1mm}M_{1,2}(y,z)\hspace{-1mm}=\hspace{-1mm}\sqrt{\frac{1+z^{2}}{2}
+y \pm\sqrt{\frac{(1-z^{2})^{2}}{4}+y(1+z^2)}}.
\ea
To study the case $\Delta_T\ll\gamma_{1}$ we change $y\rightarrow zy$ and then take the limit $z\to0$.
We find that $\sigma_{xy}=4e^{2}/h$. The numerical study of Eq.(\ref{dc-Hall}) shows that $\sigma_{xy}$ does not depend of $z$ and always equals  $4e^{2}/h$. Actually, this is a reflection of the fact that
the dc Hall conductivity $\sigma_{xy}$ can be written in terms of the topological Pontryagin
index,\cite{Tse-PRB}
\be
{\cal C}= \frac{1}{24\pi^{2}}\epsilon_{\mu\nu\lambda}\int d\omega d^{2}p\,{\rm tr}\left[
G\partial_{\mu}G^{-1}G\partial_{\nu}G^{-1}G\partial_{\lambda}G^{-1}\right],
\ee
where $\partial_{\mu}G^{-1}={\partial G^{-1}}/\partial p_{\mu}$, $p_{\mu}=(\omega,p_{x},p_{y})$,
$\epsilon_{\mu\nu\lambda}$ is the antisymmetric tensor, and $G=(i\omega -H)^{-1}$ is
the Green's function on the imaginary frequency axis.\cite{Volovik}

In the case of a more general gap $\Delta_{\xi s}=U+\xi\Delta_T$ the behavior
of the dc Hall conductivity is different: it is zero for $\Delta_{T}<U$
and $\sigma_{xy}=4e^{2}/h$ for $\Delta_{T}>U$.

We also provide the analytic expression for the complex diagonal optical conductivity
\begin{widetext}
\begin{equation}
\label{sigma-xx-eq}
\begin{split}
\sigma_{xx}(\Omega)= \frac{e^{2}}{i h\Omega}
\sum\limits_{\xi=\pm}\int\limits_{0}^{\infty}\frac{dx}{(E_{1}^{2}-E_{2}^{2})^{2}}\Bigg\{  2\gamma_{1}^{2}x
&\left[ \frac{E_{1}^{2}+\Delta_{\xi}^{2}}{E_{1}^{2}}\left(\frac{1}{2E_{1}-\Omega-2i\Gamma}
+\frac{1}{2E_{1}+\Omega+2i\Gamma} -\frac{1}{2E_{1}-2i\Gamma}-\frac{1}{2E_{1}+2i\Gamma}\right) \right.  \\
+ & \left. \frac{E_{2}^{2}+\Delta_{\xi}^{2}}{E_{2}^{2}}\left(\frac{1}{2E_{2}-\Omega-2i\Gamma}
+\frac{1}{2E_{2}+\Omega+2i\Gamma}-\frac{1}{2E_{2}-2i\Gamma}-\frac{1}{2E_{2}+2i\Gamma}\right) \right]  \\
+ \frac{1}{2E_{1}E_{2}} \left(\frac{1}{E_{1}+E_{2}-\Omega-2i\Gamma} \right. + & \left. \frac{1}{E_{1}+E_{2}
+\Omega+2i\Gamma}-\frac{1}{E_{1}+E_{2}-2i\Gamma}-\frac{1}{E_{1}+E_{2}+2i\Gamma}\right)  \\
\times \left[2E_{1}E_{2}(E_{1}^{2}-E_{2}^{2})^{2}- \right. &4\gamma_{1}^{2}E_{1}E_{2}(E_{1}^{2}
+E_{2}^{2}-\gamma_{1}^{2}) \\
 +   \Delta_{\xi}^{2}
 \left[(E_{1}^{2}+E_{2}^{2})(E_{1}-E_{2})^{2} \right. & -5\gamma_{1}^{2}(E_{1}^{2}+E_{2}^{2})   + 6\gamma_{1}^{2}E_{1}E_{2}
+4\gamma_{1}^{2}(\gamma_{1}^{2}+\Delta_{\xi}^{2})] \left.-\Delta_{\xi}^{4}(E_{1}-E_{2})^{2}\right] \Bigg\}.
\end{split}
\end{equation}
\end{widetext}
Equation~(\ref{sigma-xx-eq}) is derived from the four-band model similarly to the
optical Hall conductivity. We will use this in Sec.~\ref{sec:Faraday-Kerr}
for the analysis of Kerr and Faraday rotations.

\subsection{Optical Hall conductivity for the QAH state}
\label{four-component-C}

We immediately conclude from Eq.~(\ref{sigmaxy-clean})
that for the QVH and LAF states, $\sigma_{xy}(\Omega) =0$.
For the QSH state, the optical Hall conductivity
is also zero due to the summation over spin in Eq.~(\ref{sigmaxy-clean}).
However, it is not equal to zero  for the QAH state ($\Delta_{\xi s}=\xi\Delta_T$).
In Fig.~\ref{sigmaxy-our} the dependence of $\mbox{Re} \sigma_{xy}(\Omega)$ is plotted.
\begin{figure}[ht]
\includegraphics[width=8cm]{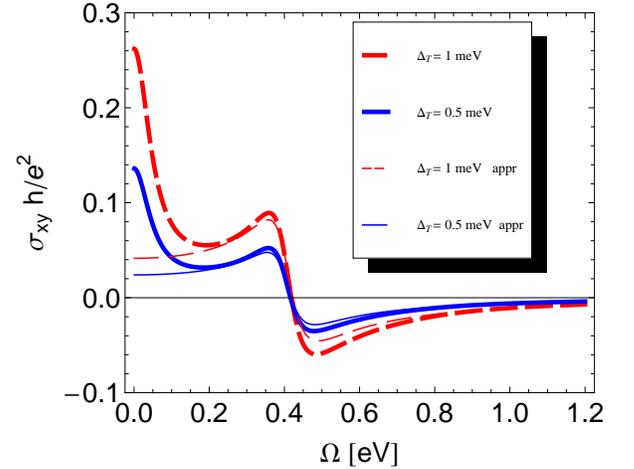}
\caption{(Color online)
The real part of the optical Hall conductivity $\sigma_{xy}(\Omega)$
in units of $e^2/h$ as a function of photon energy $\Omega$.
The thick lines are computed using Eq.~(\ref{sigmaxy-clean})
and thin lines using Eq.~(12) of Ref.~\onlinecite{Levitov2011}.
The long-dashed and solid curves are for the QAH gap
$\Delta_{T}= \SI{1}{meV}$ and $\Delta_{T}= \SI{0.5}{meV}$, respectively.  }
\label{sigmaxy-our}
\end{figure}
The thick curves are computed using Eq.~(\ref{sigmaxy-clean}) obtained in the four-band model.
For comparison with Ref.~\onlinecite{Levitov2011}
we took the same values for the parameters $\gamma_{1}= \SI{0.4}{eV}$, $2\Gamma= \SI{0.05}{eV}$
and plotted thin curves using the approximate expression (12) derived in Ref.~\onlinecite{Levitov2011}.
We also considered the two values of the gap: $\Delta_T= \SI{1}{meV}$ (long-dashed (red) curve) and
$\Delta_T= \SI{0.5}{meV}$ (solid (blue) curve). We observe that the approximate expression
is in agreement with the four-band model in the vicinity of $\Omega = \gamma_1$ and for higher energies. For low-energies
the behavior of the Hall conductivity is correctly described only by the four-band model.

In addition to the QAH gap $\Delta_T$, the gating induces
a finite time-reversal invariant gap $U_0$.
Our multigap expression (\ref{sigmaxy-clean}) for the optical Hall conductivity
also allows to study this case if one takes a gap $\Delta_{\xi s}=U+\xi\Delta_T$.
The results of the computation are presented in Fig.~\ref{sigmaxy-diffDeltas}.
\begin{figure}[ht]
\includegraphics[width=8cm]{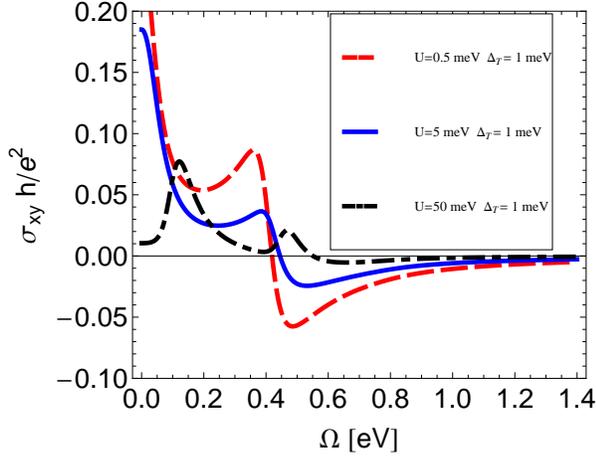}
\caption{(Color online)
The real part of the optical Hall conductivity $\sigma_{xy}(\Omega)$
in units of $e^2/h$ as a function of photon energy $\Omega$.
The QAH gap $\Delta_{T}= \SI{1}{meV}$.
The long dashed curve is for the layer asymmetry gap $U= \SI{0.5}{meV}$, the solid curve is
for $U= \SI{5}{meV}$, and the dash-dotted curve is for $U= \SI{50}{meV}$.}
\label{sigmaxy-diffDeltas}
\end{figure}
The QAH gap $\Delta_T$ is set to be $\SI{1}{meV}$ for all three curves, while the value of the layer asymmetry
gap $U_0$ is changed. The long dashed (red) curve is for $U= U_0=\SI{0.5}{meV}$, the solid (blue) curve is
for $U=U_0= \SI{5}{meV}$, and the dash-dotted (black) curve is for $U=U_0= \SI{50}{meV}$.
It is seen that in general the optical Hall conductivity is sensitive to the external electric
field $E_{\perp}$ which decreases the value of the jump of $\sigma_{xy}(\Omega)$ near $\Omega = \gamma_1$
and shifts its position to the higher energies.

\subsection{The optical Hall conductivity for $\Delta_T=0$ and superposition of three gaps}
\label{four-component-B}

The Hall conductivity ${\rm Re}\,\sigma_{xy}(\Omega)$ is
nonzero if the time-reversal symmetry is broken. This however is a necessary, but not sufficient condition.
Indeed, although the LAF state gap $U_T$ breaks time-reversal symmetry, it follows from
Eq.~(\ref{sigmaxy-clean}) that ${\rm Re}\,\sigma_{xy}(\Omega) =0$ in this state.
Then it seems that the Hall conductivity could be nonzero only when the QAH gap $\Delta_T \ne 0$.
However, the analysis performed in Sec.~\ref{sec:low-B} shows that the dc Hall
conductivity might still be nonzero even if $\Delta_T=0$. According to Eq.~(\ref{inequalities}),
this happens if three other gaps $U_T$, $U$, and $\Delta$ are
not zero and satisfy certain inequalities.

As we mentioned in Sec.~\ref{sec:model}, the two states, LAF ($\Delta_{\xi s}=sU_T$) and QSH
($\Delta_{\xi s}=\xi s\Delta)$, are the most likely candidates for the gapped ground state of
bilayer graphene at the neutral point in the absence of external fields.
The gating of bilayer induces the layer asymmetry gap $U_0$.
In the case $U_0 \ne 0$, the analysis of the mean-field gap equations shows  that the QSH state transforms
into the superposition of the QSH gap $\Delta$ and the layer asymmetry gap $U$, while the LAF state
into the  the superposition of the LAF gap $U_T$ and $U$, respectively.
Furthermore, in the QSH state with two gaps a third LAF $U_T$ gap may also open
in the presence of magnetic impurities. Similarly, there may be a mechanism to
generate a QSH gap, $\Delta$  for the LAF state with two gaps. Since these additional third gaps
naturally should be much smaller than the other two, in this subsection we will consider
the optical Hall conductivity for the two cases of superpositions of three gaps. In the first case,
$U_T$ is much smaller than $\Delta$, $U$ and, in the second,
$\Delta$ is much smaller than $U_T$, $U$. However, because of
the symmetry of the Hall conductivity under the interchange $U_{T}\leftrightarrow\Delta$,
the second case is in fact equivalent to the first one. In Fig.~\ref{fig:6} we plotted the dependence
${\rm Re}\,\sigma_{xy}(\Omega)$ for fixed gaps $U_{T}=\SI{1}{meV},\Delta=\SI{5}{meV}$ and for three different
values of the layer asymmetry gap $U=\SI{5}{meV},\SI{10}{meV},\SI{20}{meV}$ and $\Gamma=\SI{25}{meV}$.
\begin{figure}[ht]
\includegraphics[width=8cm]{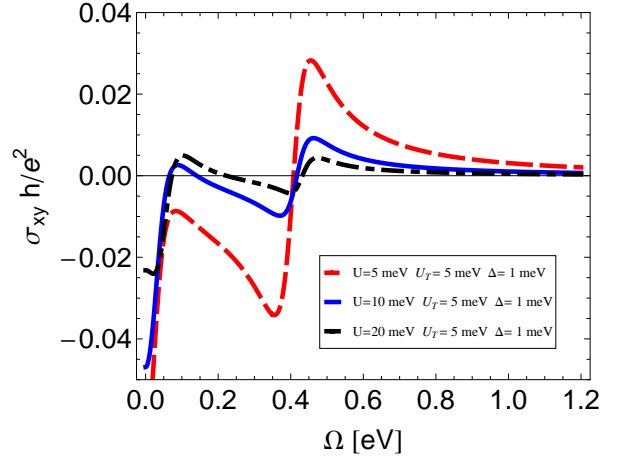}
\caption{(Color online)
The real part of the optical Hall conductivity $\sigma_{xy}(\Omega)$
in units of $e^2/h$ as a function of photon energy $\Omega$.
The QAH gap $\Delta_{T}= 0$. The LAF gap  $U_{T}=\SI{1}{meV}$, and the QSH gap  $\Delta=\SI{5}{meV}$.
The long dashed (red) curve  is for the layer asymmetry gap $U= \SI{5}{meV}$, the solid (blue) curve is
for $U= \SI{10}{meV}$, and the dash-dotted (black) curve is for $U= \SI{20}{meV}$.}
\label{fig:6}
\end{figure}
We observe that the behavior of $\mbox{Re} \sigma_{xy}(\Omega)$ becomes more rich than in pure QAH state
and in addition to the jump at $\Omega = \gamma_1$ a new kink at lower energy develops. For
smaller values of $U = \SI{5-10}{meV}$ this kink results in the
two additional zeros on the function $\sigma_{xy}(\Omega)$.

This shows that if the Faraday or Kerr rotation is observed in zero magnetic field, not only
can one state that some time-reversal symmetry state is present, but also shed a light on the specific nature of this state, by studying the dependence $\sigma_{xy}(\Omega)$. To plot the figures in
this section, we used a high value of the scattering rate $\Gamma$ which shows that the
observed features are rather robust with respect to the disorder. This allows us to conclude
that these effects can be observed experimentally even if the values of the gaps are smaller
than $\Gamma$.

\section{Relation between the optical Hall conductivity and the Faraday/Kerr rotation angles}
\label{sec:Faraday-Kerr}

Spontaneously broken time-reversal symmetry in bilayer graphene should be
observable via optical polarization rotation when light is transmitted
through the sample (Faraday effect) or reflected by it (Kerr effect).\cite{Levitov2011} If
graphene is deposited on a substrate, the rotation angles depend not only on
the optical conductivity of graphene, but also on the substrate properties. In
this section, we provide formulas and calculate Faraday and Kerr rotation
spectra for three practically relevant situations shown in the inset of
Fig.~\ref{fig:7}, namely, (i) for free-standing graphene, (ii) for graphene on
a thick substrate with a refractive index $n = 1.5$, (which closely matches
the properties of SiO$_{2}$ and boron nitride in the spectral range of interest),
and (iii) for graphene on a dielectric layer with $n=1.5$ and a thickness of 
$d = \SI{300}{nm}$ on top of a thick layer with $n_s=3.5$
(which corresponds to the most commonly used SiO$_2$/Si substrates).

If graphene is deposited on a thick substrate with refractive index $n$, the
experimental Kerr and Faraday rotation angles are defined, respectively, by
the relations:
\begin{equation}
\label{Kerr-main}
\theta_{K} = \frac{\mbox{arg}(r_{-}) - \mbox{arg}(r_{+})}{2},
\end{equation}
\begin{equation}
\label{Faraday-main}
\theta_{F} = \frac{\mbox{arg}(t_{-}) - \mbox{arg}(t_{+})}{2},
\end{equation}
where
\begin{eqnarray}
r_{\pm}  = \frac{1 - n - Z_{0}\sigma_{\pm}}{1 + n + Z_{0}\sigma_{\pm}},
t_{\pm}  = \frac{2}{1 + n + Z_{0}\sigma_{\pm}}
\end{eqnarray}
are the reflection and transmission coefficients at the
'vacuum-film-substrate' interface, $\sigma_{\pm}$ is the optical conductivity
for the right ('+') and left ('-') circularly polarized light calculated using
Eqs.~(\ref{sigma-xx-eq}) and (\ref{sigmaxy-clean}), and $Z_{0} =
4\pi/c$ is the impedance of vacuum. The case of free standing graphene is
covered by the same relations if $n$ is set to 1.

In the limit $Z_{0}|\sigma_{\pm}(\Omega)|\ll n - 1$ the formulas
(\ref{Kerr-main}) and (\ref{Faraday-main}) are greatly simplified (see also
Ref.\onlinecite{Levitov2011}):
\begin{equation}
\label{Kerr-appr}
\theta_{K} (\Omega)\approx -\frac{2Z_{0}\mbox{Re}\sigma_{xy}(\Omega)}{n^2-1},
\end{equation}
\begin{equation}
\label{Faraday-appr}
\theta_{F}(\Omega) \approx \frac{Z_{0}\mbox{Re}\sigma_{xy}(\Omega) }{n+1},
\end{equation}
from where it is obvious that both angles are proportional to the real part
of $\sigma_{xy}(\Omega)$. Note that the approximation (\ref{Kerr-appr}) is not correct for the free-standing graphene.

In Fig.~\ref{fig:7} we show the calculated Faraday and Kerr rotation spectra for free-standing graphene
(the solid line) and for graphene on top of a thick substrate with $n = 1.5$ (the dashed line),
expressed in units of the fine structure constant $\alpha$.
The exact relations (\ref{Kerr-main}) and (\ref{Faraday-main}) were used.
As an example, we take the case, where $\Delta_T = 1$ meV and other gaps are equal to zero (Fig.~\ref{sigmaxy-our}).
One can see that the Faraday angle does not differ much in the two cases and it indeed matches the real part of
the Hall conductivity. On the contrary, the Kerr rotation for the free-standing sample 
is about 10 times larger
than for supported graphene. This does not necessarily mean, however, that the Kerr geometry is favorable to
detect the gapped states in free standing graphene, since the reflection coefficient itself is proportional to $\alpha^2$
and therefore very small.\cite{StauberPRB08}

Next, we consider graphene on a double-layer substrate. In this case, Eqs.~(\ref{Kerr-main}) and (\ref{Faraday-main}) are still valid,
but the reflection and transmission coefficients are calculated differently:
\begin{equation}
\begin{split}
r_{\pm} &= r_{01\pm} + \frac{t_{01\pm}r_{12}\tau^2 t_{10\pm}}{1 - \tau^2 r_{12}r_{10\pm}},\\
t_{\pm} &= \frac{t_{01\pm}\tau t_{12}}{1 - \tau^2 r_{12}r_{10\pm}},
\end{split}
\end{equation}
\noindent where
\begin{equation}
\begin{split}
r_{01\pm} & = \frac{1 - n - Z_{0}\sigma_{\pm}}{1 + n + Z_{0}\sigma_{\pm}}, \quad
t_{01\pm} = \frac{2}{1 + n + Z_{0}\sigma_{\pm}}, \\
r_{10\pm} & = \frac{n - 1 - Z_{0}\sigma_{\pm}}{1 + n + Z_{0}\sigma_{\pm}}, \quad
t_{10\pm} = \frac{2n}{1 + n + Z_{0}\sigma_{\pm}},\\
r_{12} & = \frac{n - n_s}{n + n_s}, \quad
t_{12} = \frac{2n}{n + n_s}, \\
\tau & = \exp\left\{i(\Omega/c)nd\right\}.
\end{split}
\end{equation}

\noindent Here, $r_{ij\pm}$ and $r_{ji\pm}$ are the reflection and transmission coefficients at the interface
between media $i$ and $j$ (0,  vacuum; 1 and 2, the first and the second substrate layers) 
and $\tau$ is the transmission coefficient for the first substrate layer.

This calculation takes fully into account the Fabry-Perot interference in the 300 nm layer but not in the thick substrate.
The corresponding results  are shown by the short-dashed curves in Fig.~\ref{fig:7}. Due to the Fabry-Perot effect,
the Faraday angle and especially the Kerr angle are no longer determined by $\mbox{Re}\sigma_{xy}(\Omega)$ only.
Interestingly, the Kerr angle above 0.4 eV is even inverted with respect to the case of a simple substrate.
\begin{figure}[ht]
\includegraphics[width=8cm]{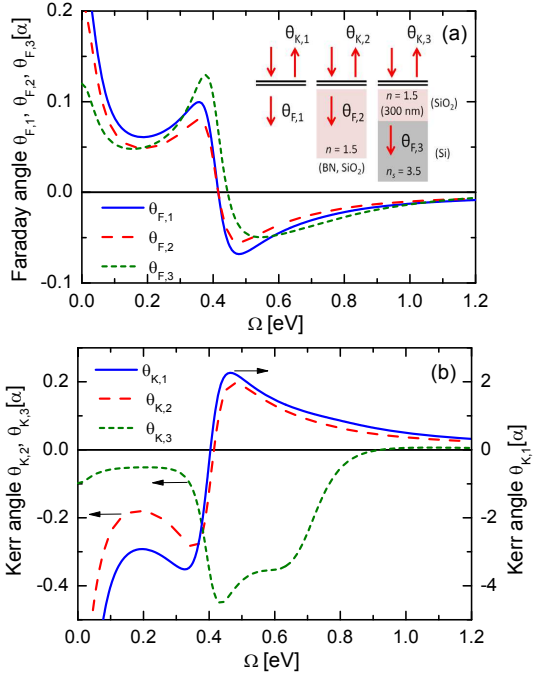}
\caption{(Color online) The calculated Faraday and the Kerr rotation angles for different experimental geometries
shown in the inset. Optical conductivity of graphene was calculated using $\Delta_T = 1$ meV and $\Delta = U = U_T = 0$.}
\label{fig:7}
\end{figure}

\section{Discussion}
\label{sec:concl}

In this paper, we studied the influence of different kinds of gaps in bilayer graphene in the two- and four-band models on
longitudinal and transverse optical conductivities paying special attention to
gaps that break the time-reversal symmetry. The two-band model is valid for energies $E<100\,\mbox{meV}$
and the four-band model is applicable up to energies when continuum approximation is valid, i.e., for wave vectors $k a\ll1$ where $a$
is the lattice constant. The upper bound $100\,\mbox{meV}$ in the two-band model is
due to neglecting the high-energy bands. Corresponding restrictions on gaps and on transverse
voltages are related to limitations of these models.

Starting from a low-energy two-band Hamiltonian,
we derived a simple analytical expression (\ref{magneto-cond-final}) for the complex magneto-optical
conductivity for two opposite circular polarizations of light. It is verified that for zero values
of the gaps, the strengths of the absorption lines satisfy the optical spectral weight conservation.
When there is a layer asymmetry gap, the corresponding absorption peak splits into two, while for
the time-reversal symmetry gap, the position of the peak shifts, but it remains unsplit.

The limit of zero magnetic field was analyzed for an arbitrary carrier density in the
two-band approximation.  We find that the necessary (but not sufficient) condition for
the optical Hall and dc Hall conductivities to remain finite is the presence of nonzero
time-reversal symmetry breaking gap. In addition to the canonical time-reversal
symmetry breaking QAH state considered in Ref.~\onlinecite{Levitov2011} which provides
$Re\,\sigma_{xy}/(e^2/h)=\pm 4$ reflecting the presence of four topologically protected
edge states for the QAH ground state of bilayer graphene, we find another more sophisticated
possibility of nonzero dc Hall conductivity $Re\,\sigma_{xy}/(e^2/h)=\pm 2$
if time-reversal symmetry breaking LAF gap, QSH gap, and QVH gap are present and satisfy
a certain inequality.

Using the full four-band model we derived analytic expressions (\ref{sigmaxy-clean}) and
(\ref{sigma-xx-eq}) for the optical Hall and longitudinal conductivities in a neutral bilayer
graphene taking into account the presence of four different gaps. We found that
the optical Hall conductivity as a function of the energy of photon is strongly sensitive to
the presence of different time-reversal symmetry breaking states. Meanwhile the real part of
the optical conductivity for the QAH state has a unique zero (see Fig.~4), the real part of the
optical conductivity for the state with a superposition of LAF, QSH, and QVH gaps may have two zeros as a function of the
energy of photon. Therefore, the observation of the optical Hall conductivity in zero magnetic field is a very
effective probe of the ground state of bilayer graphene.

The time-reversal symmetry breaking states are expected to be observed experimentally via optical
polarization rotation either in the Faraday or Kerr effects. We analyzed a possibility of such experiments
for a free standing graphene, graphene on a thick substrate and graphene on a double-layer substrate.
In the last case the Faraday angle and especially the Kerr angle are no longer determined by the real part of the
optical Hall conductivity only. Moreover, the sign of the Kerr angle is even inverted with respect
to the case of a simple
substrate.

In the this paper, we calculated optical conductivities in a very
simple approximation where quasiparticle gaps were introduced into
the Green's functions phenomenologically while the vertex
corrections due to Coulomb and other interactions were neglected.
Below the band gap the spectrum of the system may contain Coulomb bound
electron-hole pairs which would reveal themselves as poles in the
vertex function, hence as resonances in the absorption spectra.
The spectrum of resonances depends on a type of dynamically generated
gap and experimental observation of these resonances could serve as
another fingerprint for a given gapped ground state of bilayer
graphene. Detailed study of such modes is beyond of the scope of this
paper and is postponed for the future.

\begin{acknowledgments}

This work was supported by the Scientific Cooperation Between Eastern Europe
and Switzerland (SCOPES) programme under Grant
No.~IZ73Z0-128026 of the Swiss National Science Foundation (SNSF).   E.V.G.,
V.P.G. and S.G.Sh. were supported by the European FP7 program, Grant No.
SIMTECH 246937 and by the joint Ukrainian-Russian SFFR-RFBR Grant
No.~F40.2/108; A.B.K. was supported by the SNSF Grant No.~200020-130093.
V.P.G. and S.G.Sh. also acknowledge a collaborative grant from the Swedish
Institute.
\end{acknowledgments}

\appendix

\section{Diagonal conductivity spectral weight}
\label{sec:spectral-weight}

The analysis of the spectral weight proved to be useful both for theoretical (see e.g.
Refs.~\onlinecite{Gusynin2007JPCM,Benfatto2008PRB,Nicol2008PRB}) and experimental,
Ref.\onlinecite{Crassee2011PRB}, studies of graphene. Here we consider the optical spectral weight
that falls between $\Omega = 0$ and
$\Omega= \Omega_m$  with $\Omega_m$ a variable upper limit in the integral
\begin{equation}
\label{spectral-weight-def}
W(\Omega_m) = \int_0^{\Omega_m} d \Omega \mbox{Re} \sigma_{xx}(\Omega)
\end{equation}
to verify that this weight is conserved irrespectively the value of the magnetic field.
Setting $\Delta_\xi =0$ in Eq.~(\ref{AC-B=T=0}) we arrive at the expression
\begin{equation}
\label{ac-B=0}
\sigma_{xx}(\Omega)=\frac{2e^{2}}{\hbar}\left[ |\mu|\delta(\Omega)
+\frac{1}{4}
\theta\left(\Omega-2|\mu| \right)\right],
\end{equation}
which could also be   derived directly from the $2\times 2$ Hamiltonian\cite{Cserti2007PRB} for $B=0$.
Then for $\Omega_m > 2 |\mu|$
\begin{equation}
\label{weight-B=0}
W(\Omega_m) = \frac{e^2}{2 \hbar} \Omega_m
\end{equation}
which is twice as much as the spectral weight for monolayer.\cite{Gusynin2007JPCM}
Obviously, Eqs.~(\ref{ac-B=0}) and (\ref{weight-B=0}) are applicable only in the domain of
validity of the $2 \times 2$ Hamiltonian (\ref{Hamiltonian-2*2}), e.g. for $\Omega < \gamma_1$,
but this is sufficient for our purposes.

In the zero gap case, $\Delta_\xi=0$ Eq.~(\ref{magneto-cond-final}) acquires a simple form
\begin{equation}
\label{magneto-cond-D=0}
\begin{split}
 \sigma_{\pm} & (\Omega)=\frac{2 e^{2}\hbar^{3}}{\pi m^{2}l^{4}}\sum\limits_{k=0}^{\infty}
(k+1)\\
& \times \sum_{ \lambda,\lambda'=\pm}
\frac{n_{F}(\lambda M_{k+1})-n_{F}(\lambda' M_{k+2})}{\lambda'M_{k+2}-\lambda M_{k+1}}\\
& \times \frac{i}{\Omega\mp\lambda' M_{k+2}\pm\lambda M_{k+1}+2 i \Gamma}.
\end{split}
\end{equation}
Assuming for simplicity that $M_0 < \mu < M_{2}$
we get in the limit $T \to 0$ and $\Gamma \to 0$ that the diagonal optical conductivity is
\begin{equation}
\label{cond-xx}
\mbox{Re} \sigma_{xx}(\Omega) = \frac{2 e^2}{\hbar}  \omega_c^2 \sum_{k=0}^\infty (k+1)
\frac{\delta(\Omega-M_{k+2} - M_{k+1})}{M_{k+2} + M_{k+1}}.
\end{equation}
Accordingly for $\Delta_\xi=0$, the spectral weight equals
\begin{equation}
\label{weight-B}
\begin{split}
W(\Omega_m)& =\frac{2 e^2}{\hbar} \omega_c^2 \sum_{k=0}^\infty (k+1)
\frac{\theta(\Omega-M_{k+2} - M_{k+1})}{M_{k+2} + M_{k+1}} \\
& = \frac{2 e^2}{\hbar}  \omega_c \sum_{k=0}^N  \frac{\sqrt{k+1}}{\sqrt{k+2} + \sqrt{k}}\,,
\end{split}
\end{equation}
where the maximal $N$ is estimated from the condition
$\Omega_m = M_{N+2} + M_{N+1} \approx 2\omega_c N$.
The sum over $k$ in (\ref{weight-B}) can be evaluated  analytically,
\be
\sum_{k=0}^N  \frac{\sqrt{k+1}}{\sqrt{k+2} + \sqrt{k}} = \frac{1}{2}\sqrt{(N+1)(N+2))}\approx\frac{N}{2}.
\ee
Then, we again arrive at the result (\ref{weight-B=0}), obtained for $B=0$ 
confirming the conservation of the spectral weight.

To be specific, we have considered explicitly in this section only
the case $M_0< \mu < M_2$. For $\mu \in ]M_{N},M_{N+1}[$, $N\ge2$,  we can
show that the missing spectral weight in the lines $n\leq N$ is provided by the single intraband line at $M_{N+1}-M_N$.

Using a representation, similar to  Eq.~(\ref{cond-xx}), for the conductivity in the limit
$\Gamma,T \to 0$, but written for the case $\mu \in ]M_{N},M_{N+1}[$ under investigation,
we obtain the optical spectral weight lost in
units of $(e^2/\hbar)\omega_c$:
\begin{equation}
\sum_{k=0}^{N-2} \frac{(k+1)\omega_c}{M_{k+2}+M_{k+1}} + \frac{N\omega_c}{M_{N+1} + M_{N}} =
\frac{N\omega_c}{M_{N+1} - M_{N}}\,.
\end{equation}
The first term on the left hand side is the spectral weight from all
lines that disappeared from $k=0$ to $N-2$. The
second term is due to the reduction in intensity by factor $1/2$
of the line at $k=N-1$. The quantity on the right-hand side is the
optical weight of the intraband line which has picked up all of the
lost intensity. Any violation of the individual spectral weight of the lines
would violate this conservation of the optical spectral weight.

\end{document}